\documentclass[aps,nofootinbib,amsfonts,groupedaddress]{revtex4}
\usepackage{amsmath}
\usepackage{epsfig,epsf}
\usepackage{graphicx}
\usepackage{verbatim}
\usepackage{epstopdf}

\def\beq{\begin{equation}}
\def\eeq{\end{equation}}
\def\bea{\begin{eqnarray}}
\def\eea{\end{eqnarray}}
\def\eq#1{{Eq.~(\ref{#1})}}
\def\fig#1{{Fig.~\ref{#1}}}

\newcommand{\Lb}{\left(}
\newcommand{\Rb}{\right)}

\begin{document}

\voffset1.5cm
\title{Gluon saturation and energy dependence of hadron multiplicity in pp and AA collisions at the LHC }
\author{Eugene Levin$^{1,2}$ and Amir H. Rezaeian$^1$}
\affiliation{
$^1$ Departamento de F\'\i sica, Universidad T\'ecnica
Federico Santa Mar\'\i a, Avda. Espa\~na 1680,
Casilla 110-V, Valparaiso, Chile \\
$^2$ Department of Particle Physics, Tel Aviv University, Tel Aviv 69978, Israel}
 
\date{\today}
\begin{abstract}

The recent results in $\sqrt{s}=2.76$ TeV Pb+Pb collisions at the Large
Hadron Collider (LHC) reported by the ALICE collaboration shows that
the power-law energy-dependence of charged hadron multiplicity in
Pb+Pb collisions is significantly different from p+p collisions. We
show that this different energy-dependence can be explained by
inclusion of a strong angular-ordering in the gluon-decay cascade
within the Color-Glass-Condensate (or gluon saturation) approach. This
effect is more important in nucleus-nucleus collisions where the
saturation scale is larger than 1 GeV. Our prescription gives a good
description of the LHC data both in p+p and Pb+Pb collisions.

\end{abstract}
\maketitle


\section{Introduction}

The recent LHC data on hadron production in proton-proton ($pp$) and
nucleus-nucleus ($AA$) scattering \cite{CMS,AT,Al1,Apb1,Apb2}
shows that gluon saturation that follows both from the BFKL Pomeron
calculus \cite{GLR} and from the Color Glass Condensate (CGC)
approach \cite{MV1,MV2,BK,RR}, gives an adequate description of the
high energy scattering in QCD. The model based on the gluon saturation
was able to predict the hadron production at $\sqrt{s}= 7~ \text{TeV}$
\cite{LRPP} (see also Ref.~\cite{MLP}) and the experimental data both for $pp$ 
\cite{CMS} and $AA$ collisions \cite{Apb1,Apb2} confirmed the
basic qualitative predictions of this approach \cite{KLNLHC,AAM}. However, the recently
reported data from the ALICE collaboration \cite{Apb1} on hadron production in
$AA$ collisions also demonstrated that we are far away from the high
precision quantitative description. For example, the model that
predicted $7 ~\text{TeV}$ data for $pp$ scatterings and
which also describes HERA and RHIC data, failed to describe the
multiplicity in $AA$ collisions \cite{LRAA} with the
same accuracy. This fact cannot be considered as discouraging since
$AA$ collisions are more complicated QCD problem and moreover the
other model calculations, based on the same ideas,
were somehow able to describe the data \cite{KLNLHC,AAM,JA}. Nevertheless, this gives rise to a
question that whether despite of considerable progress in theory
during the past two decades we are ready to give a reliable prediction
in the framework of the high-density QCD.

In practice, our theoretical description of hadron-hadron and
nucleus-nucleus scatterings is based on two main ingredients:
Balitsky-Kovchegov (BK) non-linear equation
\cite{BK} and $k_t$ factorization \cite{KTINC,kt-rest}. However, the
BK equation is not complete since it does not take into account the
correct (non-perturbative) behaviour at large impact parameters and,
because of this, it leads to the violation of the Froissart theorem,
see Ref. \cite{KOWI} where this problem discussed in details. A
practical consequence of this is the fact that we may not be able to
guarantee the accuracy better than $\pm 20\%$ (if not worse)
\cite{BST,KOLE,BES} from the application of the BK equation. 
Having said that, the recent application of the BK equation to the
description of HERA data looks promising indeed \cite{Ja2}. On the other
hand, the $k_t$ factorization is not reliable for dense-dense system
scatterings
\cite{KRVE,BLM,LKTF} and, strictly speaking, we cannot apply this factorization
neither to proton-proton nor to nucleus-nucleus scatterings at 
midrapidity. Therefore, we are doomed to build models trying to get a
feedback from the experimental data for a theoretical breakthrough.

\begin{figure}[ht]
       \includegraphics[width=10 cm] {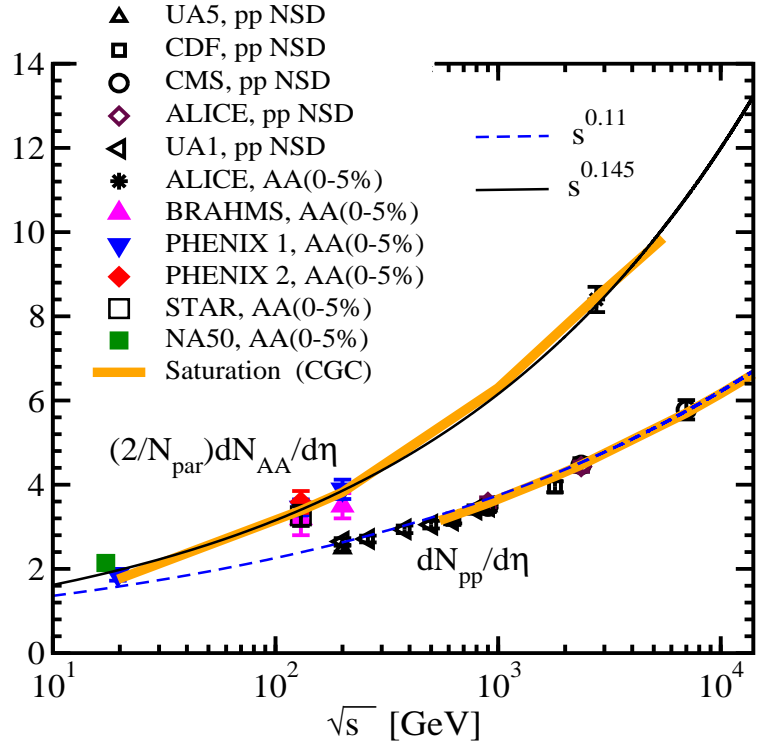}
\caption{The energy behaviour of charged particle pseudo-rapidity per participant pair for central $AA$ and non-singlet diffractive $pp$ collisions. The energy dependence can be described based on the saturation picture by $s^{0.11}$ for $pp$ and $s^{0.145}$ for $AA$ collisions. The saturation (CGC) curve for $pp$ collisions 
is taken from Ref.~\cite{LRPP}. The saturation curve for the $AA$
collisions was calculated from \eq{kt} having incorporated the effects
of gluon-jet angular ordering which is important when the saturation
scale $Q_s>1$ GeV, see the text for the details. The total theoretical
uncertainties in the saturation model calculation is about $7\%$ (not
shown here). The experimental data are from
Refs.~\cite{CMS,Al1,Apb1,pak,ua1,na,pho,bb,star,phen}. The data from the PHENIX collaboration denoted by PHENIX 1 and 2 can be found in Ref.~\cite{phen}.  }
\label{NS}
\end{figure}

As a first attempt toward understanding of the new LHC data on
nucleus-nucleus collisions, one may compare the experimental data with
the principal qualitative predictions of the gluon saturation.
The cornerstone of such predictions is the fact that the multiplicity
in $pp$ and $AA$ collisions are proportional to $Q^2_s \propto
s^{\lambda/2}$ \cite{GLR,LRAA,KLN,KTINC,kt-rest}, where $Q_s$ is the saturation scale, $s$ is the
center-of-mass energy squared per nucleon pair and $\lambda$ is free
parameter to be fixed with other experiments like DIS at HERA. This indicates
that the energy dependence of multiplicity in both $pp$ and $AA$
collisions should be the same, assuming that the atomic number or $A$
dependence of the saturation scale is factorizable from energy. This
simple property is in accordance with RHIC data \cite{LRAA,KLN,rhic-da}.  However, the new
ALICE data shows that multiplicities in $pp$ and $AA$ collisions have
a different energy power-law behaviour (see
\fig{NS}). 
Thus, at first sight, it looks as if that one of the principal
feature of high-density QCD is violated. In this paper, we will
argue that indeed the recent LHC data on $AA$ collisions at
$\sqrt{s}=2.76$ TeV has already opened up a new QCD regime which requires
further theoretical understanding than previously thought. Here, we
shall give a simple explanation of the different energy behaviour of
the hadron multiplicity in $AA$ and $pp$ data at high energy based on
the gluon saturation (or the CGC) scenario.

In the CGC approach, the hadron production goes in two stages:
production of gluons and subsequently
the decay of gluon-jet (or mini-jet) into hadrons. Therefore, the
multiplicity of the produced hadrons at pseudo-rapidity $\eta$ can be
calculated as a convolution of these two stages,

 \bea \label{I1}
\frac{d N_h}{d \eta \,d^2 p_T}\,\,&\propto &\,\, \frac{d N^{Gluon}}{d y \,d^2 p_T} \otimes N^{Gluon}_h( E_{jet}), \\
\frac{d N_h}{d \eta }&\propto & \, \sigma_s Q^2_s \times N^{Gluon}_h\Lb Q_s\Rb,   \label{I11}\
\eea
where the first part $\frac{d N^{Gluon}}{d yd^2 p_T}$ gives the gluon jet
production yield at rapidity $y$ (in $pp$ or $AA$ collisions)
computable in the $k_T$ factorization scheme \cite{KTINC,kt-rest}, and the second term
$N^{Gluon}_h$ is the average multiplicity of hadrons in the gluon jet
with a jet energy $E_{jet}$ (see Secs. II, III). The symbol $\otimes$ indicates a
convolution, that is, integrals over variables with possible weight factors included. 

 The kinematics looks simpler in the center-of-mass of the produced
 gluon in which two gluons with the mean transverse momenta of the
 order of $Q_s$ and the fraction of energy $x_1=x_2 = Q_s/\sqrt{s}$
 collide, producing the gluon which moves in the transverse plane with
 the value of its momentum of the order of $Q_s$. Eq.~(\ref{I11}) up to
 a possible logarithmic correction, can be simply obtained by a
 dimensionality argument based on the CGC picture in which the
 multiplicity of the gluon jets is proportional to $ \sigma_s Q^2_s$ where
 $\sigma_s$ is the effective area of interaction \cite{KLN}. Notice that the typical
 transverse momentum in \eq{I1} is of the order of the saturation
 scale $Q_s$.

\begin{figure}[t]
       \includegraphics[width=17 cm] {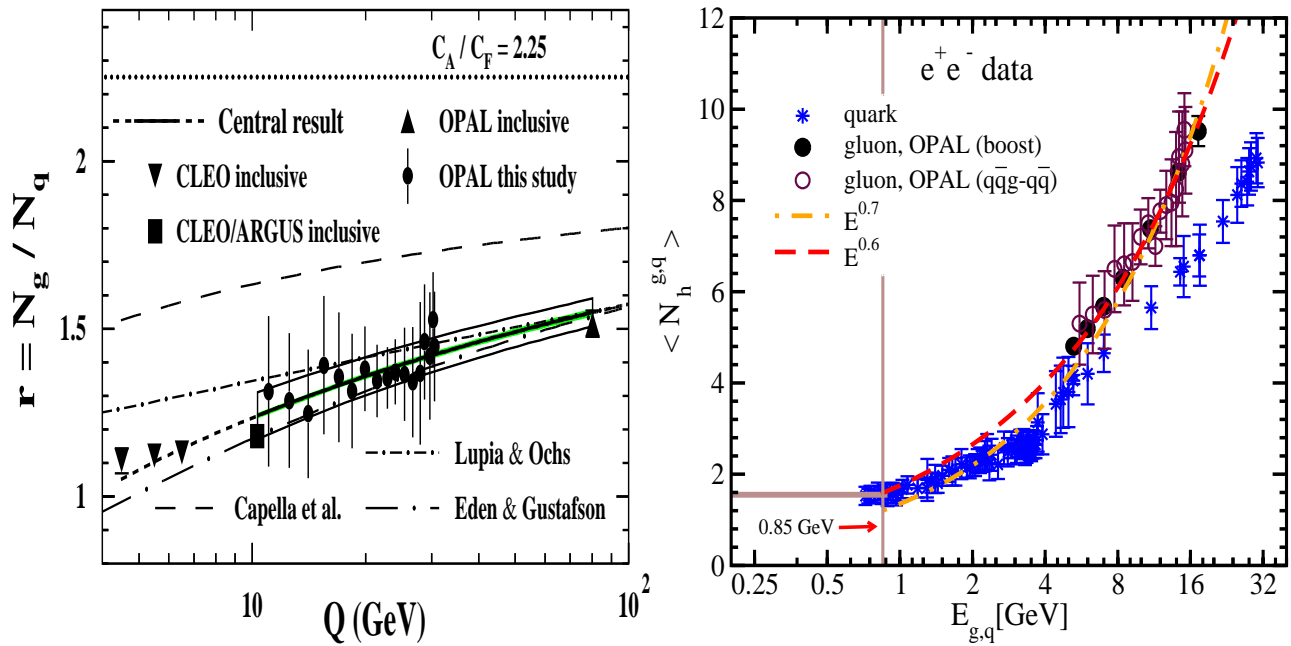}
\caption{Right: The mean charged hadron multiplicity of unbiased gluon $N^g_h$ and quark $N^q_h$ jets in $e^+e^-$ annihilation, as a function of the jet energy . For gluon jet we show experimental data obtained by two different methods, jet boost algorithm and subtracting multiplicities in two-jet $q\bar q$ events from three-jet $q\bar qg$ events \cite{ee,ee-plot}. The experimental data for quark jet production in $e^+e^-$ annihilation are taken from Ref.~\cite{pak}. The energy behaviour of  $N^{g}_h$ can be described by $E_{g}^{0.6\div 0.7}$ for $E_{g}\geq 0.85$ GeV. Left: The ratio of the mean charged particle multiplicities between unbiased gluon and quark jets as a function of scale. Various theoretical predictions \cite{pee} based on perturbative QCD (pQCD) are also shown in the plot. The plot in the left panel is taken from Ref.~\cite{ee-plot}. }
\label{mult}
\end{figure}

The crucial ingredient which is essential to explain
the different energy dependence of the multiplicity in $pp$ from $AA$
collisions originates from the experimental data for jet production in
$e^+e^-$ annihilation
\cite{ee,ee-plot,pak,LPHD}, namely $N^{Gluon}_h$ is almost constant at energies of the
gluon-jet less than about $1$ GeV but starts to increase with the energy of the
gluon-jet larger than $1$ GeV, see \fig{mult} and Sec. III.  For proton-proton collisions, the value of the saturation scale is smaller than $1$ GeV and consequently $N^{Gluon}_h$ does not give an additional
energy dependence, while for nucleus-nucleus scatterings at high
energies the saturation scale $Q_s\Lb A\Rb\,\propto A^{1/3} \, Q_s
\Lb p\Rb $ ($A$ is the atomic number) is larger than $1$ GeV and
$N^{Gluon}_h$ increases leading to an additional non-negligible
power-law energy dependence. This extra contribution accounts for the gluon-decay effect before hadronization and
is missing in the $k_T$ factorization.

\begin{figure}[t]
               \includegraphics[width=14 cm] {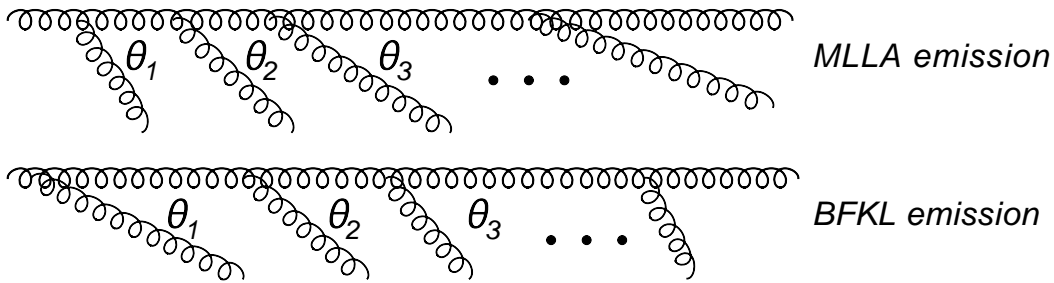}
\caption{Angular ordering in the gluon cascade in the MLLA ($\theta_1>\theta_2>\theta_3>...>\theta_n$) and the BFKL ($\theta_1<\theta_2<\theta_3<...<\theta_n$) regime. }
\label{jd2}
\end{figure}

The paper is organized as follows: In Sec. II, we introduce the
missing gluon-decay cascade effect in the $k_t$ factorization. We also
show that the observed power-law energy dependence of multiplicity in
$pp$ and $AA$ collisions at the LHC is fully consistent with the
saturation picture by inclusion of the gluon cascade angular-ordering
effect.  In Sec III, we generalize the $k_t$ factorization in order to
incorporate this effect, and present our numerical results for the
charged hadron multiplicity both in $pp$ and $pA$ collisions. As a
conclusion, in Sec. IV we highlight our main results.
\begin{boldmath}
\newline
\section{The energy-dependence of charged hadron multiplicity}
\end{boldmath}
The $k_T$ factorization \cite{KTINC,kt-rest,JMKINC,CMINC,KLINC,LPINC}
includes gluon emissions between the projectile and target, and also
gluon radiation in the {\em final} initial-state from the produced
gluons. The $k_T$ factorization accounts for the BFKL type gluon
emissions, namely the parent gluon emits a cascade of gluons with
their longitudinal momenta $k^+_i$ being progressively smaller while the
transverse momenta $k_{Ti}$ of the parent and emitted gluons are the same. This
leads to an angular ordering in the cascade shown in \fig{jd2}
\bea
&&p^+>k_1^+>k_2^+>...>k_n^+, \nonumber\\
&&p_{T}\sim k_{T1}\sim k_{T2}...\sim k_{Tn},\nonumber\\
&&\theta_1<\theta_2<\theta_3<...<\theta_n. \label{bf} \
\eea
 However, the other
contribution of the gluon decay, in the final initial-state, before
hadronization, stems from the kinematic region outside the BFKL
emission regime where both emitted gluons are collinear to the
emitter. In this kinematic region, the angle between the gluon (quark)
and the decay gluon $\theta_i$ is small and the main contribution of
gluon-decay has an opposite angular ordering to the BFKL type gluon
emissions given in Eq.~(\ref{bf}), see \fig{jd2},
\bea \label{ML}
&&p^+>k_1^+>k_2^+>...>k_n^+, \nonumber\\
&&p_{T}>> k_{T1}>> k_{T2}...>> k_{Tn},\nonumber\\
&& \theta_1>\theta_2>\theta_3>...>\theta_n. \
\eea
This angular ordering means that a gluon in a fully developed cascade
can only emit inside a cone defined by the momenta of its first two
immediate predecessors, similar to the well-known Chudakov effect in
QED \cite{QED}. It is well-established fact from jet observables (especially at
small momentum fractions $z=P_{hadron}/E_{jet}$) that soft and
collinear logarithms summed by the Modified Leading Logarithmic
Approximation (MLLA), together with angular ordering reproduces the
most important features of QCD cascade \cite{ee,ee-plot,LPHD,MDLA}. This combined with the Local
Parton-Hadron duality (LPHD) also gives quantitative predictions for
hadron multiplicity and spectra in $e^+e^-$ and $ep$ collisions over
the whole momentum range down to momenta of a few hundred MeV \cite{LPHD,MDLA}. The
MLLA contains systematically next-to-leading logarithmic
corrections and incorporates single and double-logarithmic effects in
the development of parton cascades \cite{LPHD,MDLA}.

   One should note that although the MLLA angular-ordering kinematic
   region is quite important at the gluon (quark) decay stage, but it
   does not lead to the large double log contribution in the total
   cross-section and it contributes to the self-energy of the quark
   and, therefore, to the running QCD coupling in the case that we
   integrate over all produced gluons. Therefore, this kinematic
   region is not included in the $k_T$ factorization formula and has
   to be considered separately, namely, summing of these double log
   for the gluon jet decay leads to the extra term $N^{Gluon}_h$ in
   \eq{I1} which can be calculated within the MLLA approach. On the
   other hand, the effect of propagation and interaction of the
   produced jet in the gluonic medium with the BFKL angular-ordering
   and its saturation effect have been already taken into account in
   the $k_T$ factorization.  It is well-known that the gluon
   decay probability can be factorized from the rest of
   cross-section in $e^+e^-\to q\bar q g$ reaction \cite{MDLA}. This
   is the essence of the factorization given in \eq{I1}. Therefore,
   one may extract information about the gluon-decay stage in the MLLA
   region from gluon jet data in $e^+e^-$ collisions.

In order to verify \eq{I11}, we need to know $N^{Gluon}_h$. As we
already mentioned, one may calculate the charged hadron multiplicities
in the gluon jet $N^{Gluon}_h \Lb E_{jet}\Rb$ in pQCD within the MLLA
scheme \cite{LPHD,MDLA}. 
In order to obtain the energy dependence of the function $N^{Gluon}_h$, we use
directly experimental data for $N^{Gluon}_h \Lb E_{jet}\Rb$ in the
$e^+ e^-$ annihilation. Unfortunately, such data are limited to high gluon-jet energy $E_{jet} > 5$ GeV
\cite{ee,ee-plot}, see \fig{mult}.  For lower energy $E_{jet} < 5$ GeV, we
construct the hadron multiplicity of gluon-jet from the corresponding
multiplicity of quark-jet $N^{Quark}_h$ where we have experimental
data \cite{pak}. It is well-known that in the double log
approximation, the ratio of the multiplicities in quark and gluon jets
is equal to $N^{Gluon}_h/N^{Quark}_h = C_A/C_F=9/2$
\cite{LPHD,MDLA}. However, the higher order perturbative corrections
significantly suppress this ratio at low energies of the jet making it
close to one \cite{pee,ee-plot,3nl,OCHS}, see \fig{mult} (left panel).
One can see from
\fig{mult} that $N^{Gluon}_h $ is constant at about $E_{jet} < 1$ GeV and
it grows as a power of $E_{jet}$ at higher energies.  From the
available $e^+e^-$ collisions data shown in \fig{mult}, we found that the energy-dependence
of the mean charged particle multiplicity of gluon-jet can be approximately described by 
\beq \label{eee}
 \langle N^{Gluon}_h\rangle \,\propto \,E^{\delta}_{jet},~~~\text{with}~~\delta=0.6\div 0.7 ~~~\text{for}~~~ E_{jet}\geq 0.85 \div 1~ \text{GeV}. 
\eeq
It is essential to stress again that such behaviour
also follows from the theoretical estimates in the next-to-next-to-next-to-leading order (3NLO) pQCD \cite{pee,3nl,OCHS} in the MLLA scheme  \cite{LPHD,MDLA}.

 It is well-known that the saturation scale has the
  following energy (or $x$) behaviour \cite{GLR,MV1,MV2,BK,BALE,MUTR,MUPE,b-cgc0}
\beq \label{sat}
Q^2_s(x)\,\,=\,\,Q^2_0\,\Big(\frac{x_0}{x}\Big)^\lambda \propto\,\,\,s^{\lambda/2},
\eeq
where the saturation scale $Q_0$ is fixed at an initial value $x_0$. We assume that the typical energy
  of the gluon jet $E_{jet}$ is of the order of average
  saturation scale. Now, using \eq{I11} and  Eqs.~(\ref{eee},\ref{sat}) we obtain, 
\bea 
  \frac{dN_h}{d \eta}\Lb p p \Rb\,\,&\propto&
  \,\,Q^2_s\,\,\propto\,\,s^{\lambda/2} \,\,=\,\,s^{0.11}\,,
  \label{N21}\\ \frac{dN_h}{d \eta}\Lb A A \Rb\,\,&\propto& \,\,Q^2_s
  \times \Lb E_{jet}\,\propto \,Q_s\Rb^{0.65}\,\,\propto
  \,\,s^{\lambda/2+ 0.65\times\lambda/4}\,\,=\,\,s^{0.145},\label{N22}
  \eea where we assumed that the saturation scale for $pp$ collisions
  is $Q_s<1$ GeV and for $AA$ collisions we have $Q_s>1$ GeV. In the
  above and the following we take for the parameter $\delta$, the
  average value $\overline \delta =0.65$ from \eq{eee} and
  \fig{mult}. In \eq{N21}, the average value of $\lambda=0.11$ in the
  the effective saturation scale for $pp$ collisions can be obtained
  from $k_t$ factorization results given in Ref.~\cite{LRPP} or by a fit to the
  available data for non-singlet diffractive inclusive hadron
  production in $pp$ collisions shown in \fig{NS}. Then, the power-law
  behaviour given in \eq{N22} for $AA$ collisions comes naturally
  without any extra freedom. In \fig{NS}, we show that the energy
  power-law scaling given in Eqs.~(\ref{N21}.\ref{N22}) leads to a
  very good description of experimental data both in $pp$ and $AA$
  collisions, including the recent ALICE data in $AA$ collisions at
  $2.76$ TeV.  
 \begin{boldmath}
\newline
\section{Charged hadron multiplicity in the improved $k_t$ factorization}
\end{boldmath}
In this section, we shall investigate how the saturation model
predictions based on the $k_t$ factorization \cite{KTINC,kt-rest} will change by the
inclusion of the angular ordering effect in the gluon-jet decay
cascade. Motivated by previous sections, we postulate that the missing effect of
gluon-jet decay cascade can be effectively incorporated into the $k_t$
factorization in the following way,
\beq \label{kt}
\frac{d N_h}{d \eta}\left( AA~\text{or}~pp \right)=\,\frac{\mathcal{C}}{\sigma_{s}}
\int d^2 p_T \,h[\eta]~ \frac{d \sigma^{Gluon}}{d y \,d^2 p_{T}}\left( AA~\text{or}~ pp \right)~\mathcal{N}^{Gluon}_h(\overline Q_s),
\eeq
where $h[\eta]$ is the Jacobin transformation between $y$ and
$\eta$ \cite{KLN}. The impact-parameter dependence of the formulation allows to
calculate the average area of interaction $\sigma_{s}$ via the 
geometrical scaling property \cite{LRPP}. The gluon jet cross-section in $AA$ (or $pp$)
collisions can be obtained from \cite{KTINC}
\bea \label{M4}
&&\frac{d \sigma^{Gluon}(y;p_T;\overline{B})}{d y \,d^2 p_{T}} =
\frac{2C_F\alpha_s(p_T)}{ (2\pi)^4} \int_{B_1}^{B2} d^2 \vec B \int
d^2 \vec b d^2 \vec r_T e^{i \vec{p}_T\cdot \vec{r}_T} \frac{ \nabla^2_T N^G_{A,p}\Lb x_1; r_T; b\Rb
\nabla^2_T N^G_{A,p}\Lb x_2; r_T; b_{-} \Rb
}{p^2_T\alpha_s\left(Q_{A,p}\left(x_1;b\right)\right)\alpha_s\left(Q_{A,p}\left(x_2;b_{-}\right)\right)
}, \
\eea 
where $x_{1,2}=(p_T/\sqrt{s})e^{\pm y}$, $p_T$ and $y$ are the
transverse-momentum and rapidity of the produced gluon
jet. The vector $\vec B$ is the impact
parameter between the center of two nuclei (or two hadrons in the case
of $pp$ collisions), $\vec b$ and $\vec{b}_{-}=\vec b-\vec{B}$ are the
impact parameter between the interacting nucleons with respect to the
center of two nuclei (or hadrons). A given centrality bin corresponds
to a range of the impact-parameter $\overline{B}\in[B_1,B_2]$ of the
collisions. We extended the $k_T$-factorization by introducing a
running strong-coupling $\alpha_s$~\cite{LRPP}. In the above, the
amplitude $N^G_{A,p}$ is defined as \cite{KTINC},
\beq \label{M3}
N^G_{A,p}\Lb x_i; r_T; b \Rb =2 N_{A,p}\Lb x_i; r_T; b \Rb - N^2_{A,p}\Lb x_i; r_T; b \Rb,
\eeq
where $N_{A,p}\Lb x_i; r_T; b \Rb$ is the dipole-nucleus (for index
$A$) or dipole-proton (for index $p$) forward scattering amplitude
with $r_T$ and $\vec b$ being the transverse dipole size and the
impact parameter of the scattering, respectively.

In \eq{kt}, $\mathcal{N}^{Gluon}_h$ is the average hadron
multiplicity in the gluon-jet decay in the MLLA region and can be obtained from experimental
data in $e^+e^-$ reactions \cite{ee,ee-plot,pak}. Using Eq.~(\ref{eee}) and assuming that
typical transverse momentum of the gluon-jet is approximately equal to the average saturation
scale $\overline Q_{A,p}$ at a given centrality and kinematics, we define
\bea\label{c0}
\mathcal{N}^{Gluon}_h(\overline Q_{A,p})=C_0\left\{\begin{array}{l} \left(\frac{\overline Q_{A,p}}{0.85}\right)^{0.65} \,\,\,\,\,\mbox{for}~~~~\overline Q_{A,p} \geq 0.85~\text{GeV};\\
\\ 1 \,\,\,\,\,\,\,\,\,\,\mbox{for}~~~~\overline Q_{A,p} < 0.85   ,\end{array}
\right.
\eea
with a notation, 
\beq 
\overline Q_{A,p}=\left(\frac{Q_{A,p}^2\left(x_1,b\right)+Q_{A,p}^2\left(x_2,b_{-}\right)}{2}\right)^{1/2},
\eeq
The normalization factor $C_0$ in \eq{c0} can be absorbed into the parameter $\mathcal{C}$ in \eq{kt} which relates the produced
gluons to the final-state hadrons based on the Local Patron-Hadron
Duality principle \cite{LPHD}, assuming that the final-state hadronization is a soft
process and cannot change the direction of the emitted gluon-jet
further. 

The impact-parameter dependence in the $k_t$ factorization is not
trivial and a prior is not obvious if it can be factorized.  Here we
are interested to study the effect of new $\mathcal{N}_h^{Gluon}$ term
in the $k_t$ factorization \eq{kt}. To this end, we employ the b-CGC
saturation model \cite{b-cgc} which gives a good description of inclusive hadron
production in $pp$ collisions at the LHC \cite{LRPP}. In this model the size of proton naturally changes with energy \cite{LRPP}. 
This model effectively incorporates all known saturation properties driven by the small-x non-linear
evolution equations \cite{b-cgc0} including the impact-parameter dependence of the dipole amplitude \cite{bb-cgc}. This model describes
very well the HERA DIS data at small-x \cite{b-cgc0,b-cgc,b-cgc1} and direct-photon production \cite{me}. The extension of this model for the case of nuclear target was introduced in Ref.~\cite{LRAA} which also gives a
good description of RHIC multiplicity data.  
The dipole-nucleon forward scattering amplitude in the b-CGC model \cite{b-cgc} is defined as, 
\bea \label{M5}
N_p\Lb x; r; b\Rb=\left\{\begin{array}{l} N_0\,\Lb
\frac{\mathcal{Z} }{2}\Rb^{2 (\gamma_s\,\,+\,\,\frac{1}{\kappa \lambda
Y}\ln\Lb\frac{2}{\mathcal{Z}
}\Rb)}\,\,\,\,\,\mbox{for}\,\,\mathcal{Z}\leq 2;\\
\\ 1-\exp\Lb -\mathcal{A} \ln^2\Lb \mathcal{B}
\mathcal{Z}\Rb\Rb\,\,\,\,\,\,\,\,\,\,\mbox{for}\,\,\mathcal{Z} >2,\end{array}
\right.
\eea
where we defined $\mathcal{Z}=r\, Q_p(x;b)$, $Y=\ln(1/x)$ and $\kappa
= \chi''(\gamma_s)/\chi'(\gamma_s)$ where $\chi$ is the LO BFKL
characteristic function.  The parameters $\mathcal{A}$ and
$\mathcal{B}$ are determined uniquely from the matching of $N_p$ and
its logarithmic derivatives at $\mathcal{Z}=2$. The proton saturation scale is given by
\beq \label{M6}
Q_{p}(x;b')\,\,=\,\,\Lb \frac{x_0}{x}\Rb^{\frac{\lambda}{2}}\,\exp\left\{- \frac{b'^2}{4 (1 - \gamma_{cr}) B_{CGC}}\right\}.
\eeq
Based on the universality of the saturation in the CGC framework, the corresponding dipole-nucleus dipole amplitude $N_A$ can be obtained from \eq{M5} by only replacing the proton saturation scale by that of the nucleus,   
\beq \label{M8}
Q^2_{A}\Lb x;b \Rb\,\,\,= \,\,\, \int d^2 \vec b'~T_A\Lb\vec{b} - \vec{b}' \Rb\,Q^2_p\Lb x; b'\Rb,
\eeq
where $T_A\Lb B\Rb$ denotes the nuclear thickness. The above definition
leads to $Q_A^2\approx Q_p^2 A^{1/3}$ which is consistent with basic
idea of saturation \cite{GLR,MV1,QSA}.  We use
for the nuclear thickness the Wood-Saxon parametrization
\cite{WS}. The parameters $ \lambda, \gamma_{cr}, N_0$, $x_0$ and
$B_{CGC}$ are obtained from a fit to the DIS data at low Bjorken-$x$
$x<0.01$ with a very good $\chi^2/\text{d.o.f.}=0.92$ \cite{b-cgc}.

\subsection{Numerical results and discussion}
\begin{figure}[t]
       \includegraphics[width=9 cm] {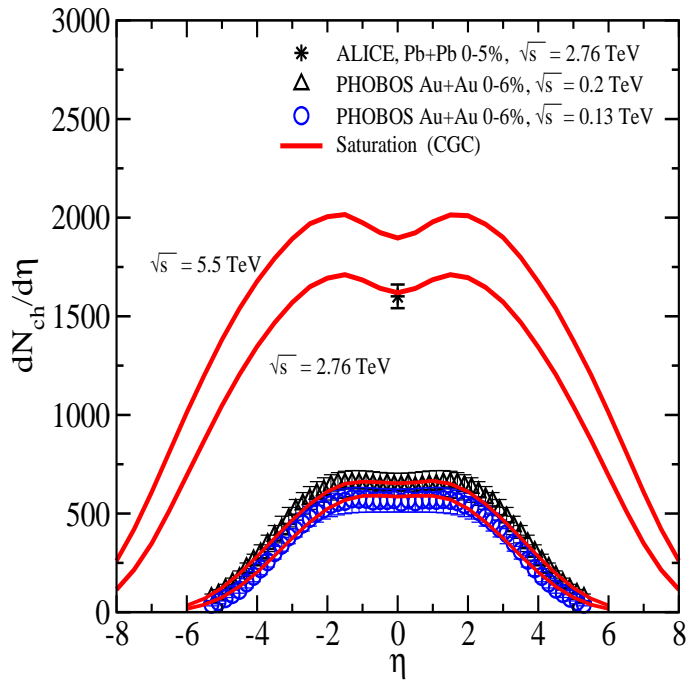}
\caption{Pseudo-rapidity distribution of charged particles
             produced in Au-Au and Pb-Pb central collisions at RHIC
             $\sqrt{s}=130, 200$ GeV and the LHC energies
             $\sqrt{s}=2.75, 5.5$ TeV. The experimental data are from
             the PHOBOS \cite{rhic1} and the ALICE collaboration \cite{Apb1}. }
\label{NS2}
\end{figure}

The number and density of participant at different centralities are
calculated based on the Glauber formalism \cite{glau} assuming
$\sigma_{nn}^{inel}=64.8, 58.5$ and $42$ mb for $\sqrt{s}=5.5, 2.76$
and $0.2$ TeV, respectively \cite{sigma}. Following
Refs.~\cite{LRPP,LRAA} in order to regularize the divergence of the
$k_t$ factorization, we introduce a gluon-jet mass $m_{jet}$. We are
now ready to confront the improved $k_t$ factorization \eq{kt} with
experimental data.  First, notice that the nuclear saturation scale
defined via \eq{M8} can be in principle different with exact one upto
a factor of the order of one. A change of $Q_A\to 1\div 1.5~Q_A$
brings about $0-7\%$ increase in the hadron multiplicity obtained from
\eq{kt} at high energies.  We have only two free parameters, the
pre-factor $\mathcal{C}$ and the gluon jet-mass $m_{jet}$ which are
determined at low-energy for a fixed centrality. The main source
of uncertainties in our approach is due to the assumption that gluon-jet mas $m_{jet}$ and the normalization pre-factor $\mathcal{C}$ do not change with energy, rapidity and centrality. Unfortunately, due to
limited available data in $AA$ collisions at various high energies we
cannot verify if this assumption is correct and therefore we should
take into account possible uncertainties associated with this
assumption.  We use RHIC data at $\sqrt{s}=200$ GeV around midrapidity
to fix our only free parameters $m_{jet}$ and
$\mathcal{C}$. Unfortunately, as it is obvious from Figs.~\ref{NS} and \ref{NS2}, the experimental errors in the data points
taken for fixing these unknown parameters is rather large. We checked
that in the case of $AA$ collisions, $m_{jet}\approx 0.12\div 0.14$ GeV is consistent with RHIC data
within error bars. The experimental errors in the charged hadron multiplicity in Au+Au
collisions at $\sqrt{s}=200$ GeV (shown in \fig{NS}) may induce an uncertainty as large as
$7\div 9\%$ in the value of the parameter $\mathcal{C}$.

In \fig{NS}, we show the energy dependence of the charged hadron multiplicity at midrapidity (labeled with saturation) obtained from
\eq{kt} both for $pp$ and $AA$ collisions. The proton saturation scale $Q_p$ in the b-CGC model \eq{M6} is rather
small and varies very slowly with energy, e. g. for central
collisions and midrapidity at $\sqrt{s}=14$ TeV and $p_T=1$ GeV, we
have $Q_p=0.8$ GeV. Therefore, in the case of $pp$ collisions for our
interested range of energy considered in this paper, the contribution
of $\mathcal{N}^{Gluon}_h$ term in the improved $k_t$ factorization
\eq{kt} is negligible and the charge hadron multiplicity obtained via
\eq{kt}, shown in \fig{NS}, coincides with the results given in Ref.~\cite{LRPP}
without the presence of $\mathcal{N}^{Gluon}_h$ term. However, in
the case of $AA$ collisions, the nuclear saturation scale defined by
\eq{M8} can be $Q_A>0.85$ GeV and consequently $\mathcal{N}^{Gluon}_h$ term in the improved $k_t$
factorization \eq{kt} is important. The inclusion of gluon-decay
angular-ordering effect via $\mathcal{N}^{Gluon}_h$ in \eq{kt} does
not noticeably affect our prescription at RHIC due to our freedom in
fitting $m_{jet}$ and $\mathcal{C}$ parameters to the same data (at
$\sqrt{s}=200$ GeV) while it increases the charged hadron multiplicity
about $20-25\%$ at the LHC energies in $AA$ collisions. Notice that
the impact-parameter dependence of condition given in \eq{c0} limits
the contribution of $\mathcal{N}^{Gluon}_h$ term at various energies
and centralities.  Overall, the improved $k_t$ factorization results
\eq{kt} shown in
\fig{NS} agree very well with both $pp$ and $AA$ data at the LHC and
also RHIC, including the recent ALICE data for $AA$ collisions at
$2.76$ TeV.

In Fig.~\ref{NS2}, we show pseudo-rapidity dependence at RHIC energies
$\sqrt{s}=130$ and $200$ GeV in  $0-6\%$ Au+Au collisions, and also for the LHC
energies $\sqrt{s}=2.76$ and $5.5$ TeV in $0-5\%$ Pb+Pb collisions. In our calculation, the number of participant at $5.5$
TeV for $0-5\%$ centrality is approximately $N_{par}=385$. Our
prediction for $dN_{AA}/d\eta$ obtained from \eq{kt} at midrapidity
for $0-5\%$ Pb+Pb collisions at $5.5$ TeV is $1897\pm 133$.
It is seen in Fig.~\ref{NS2} that as the energy increases the peak of rapidity distribution at forward (backward) becomes more pronounced due to the saturation effect.
This effect has been also observed in Refs.~\cite{LRPP,me2} in the case of $pp$ collisions.  
\begin{figure}[t]
              \includegraphics[width=8 cm] {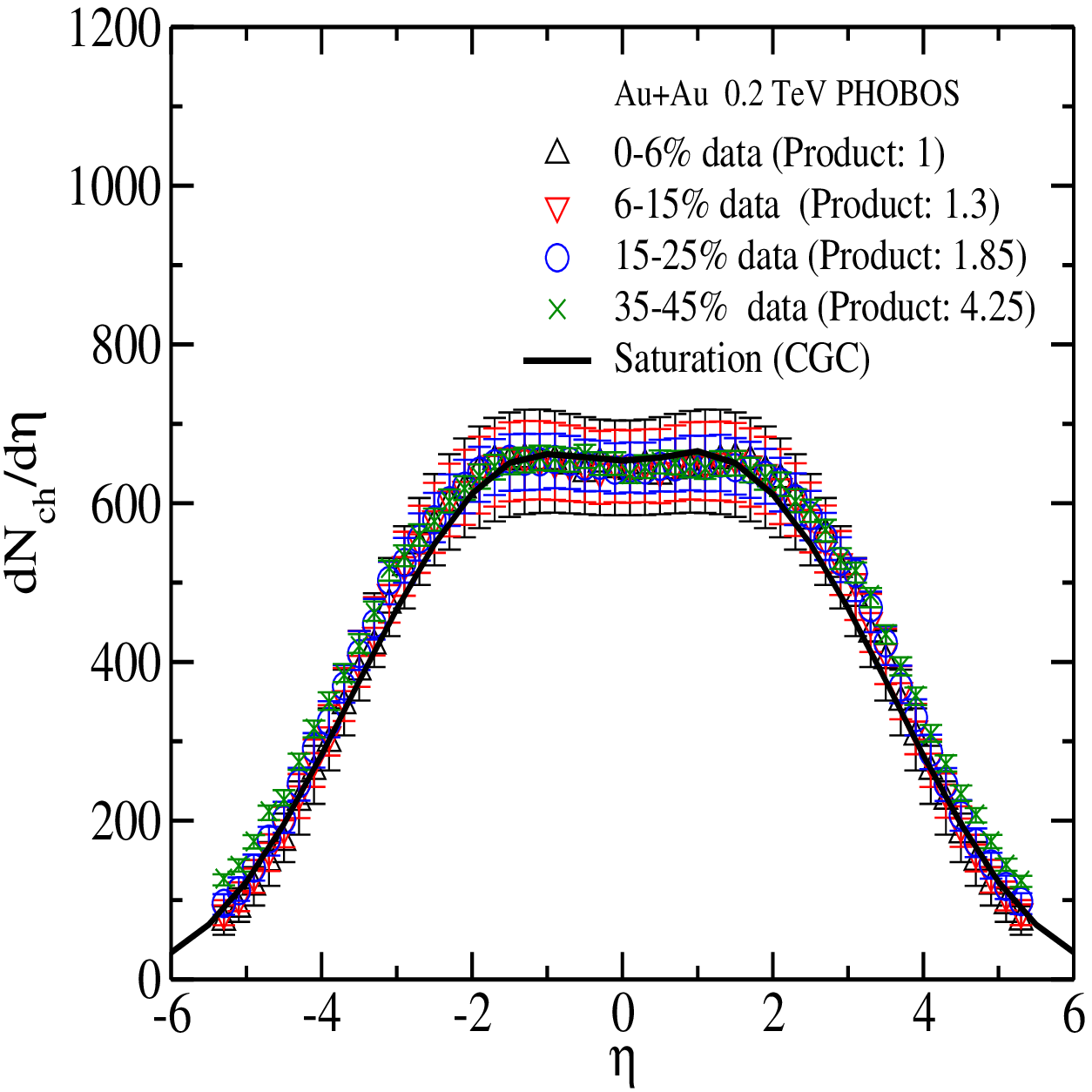}
               \includegraphics[width=8 cm] {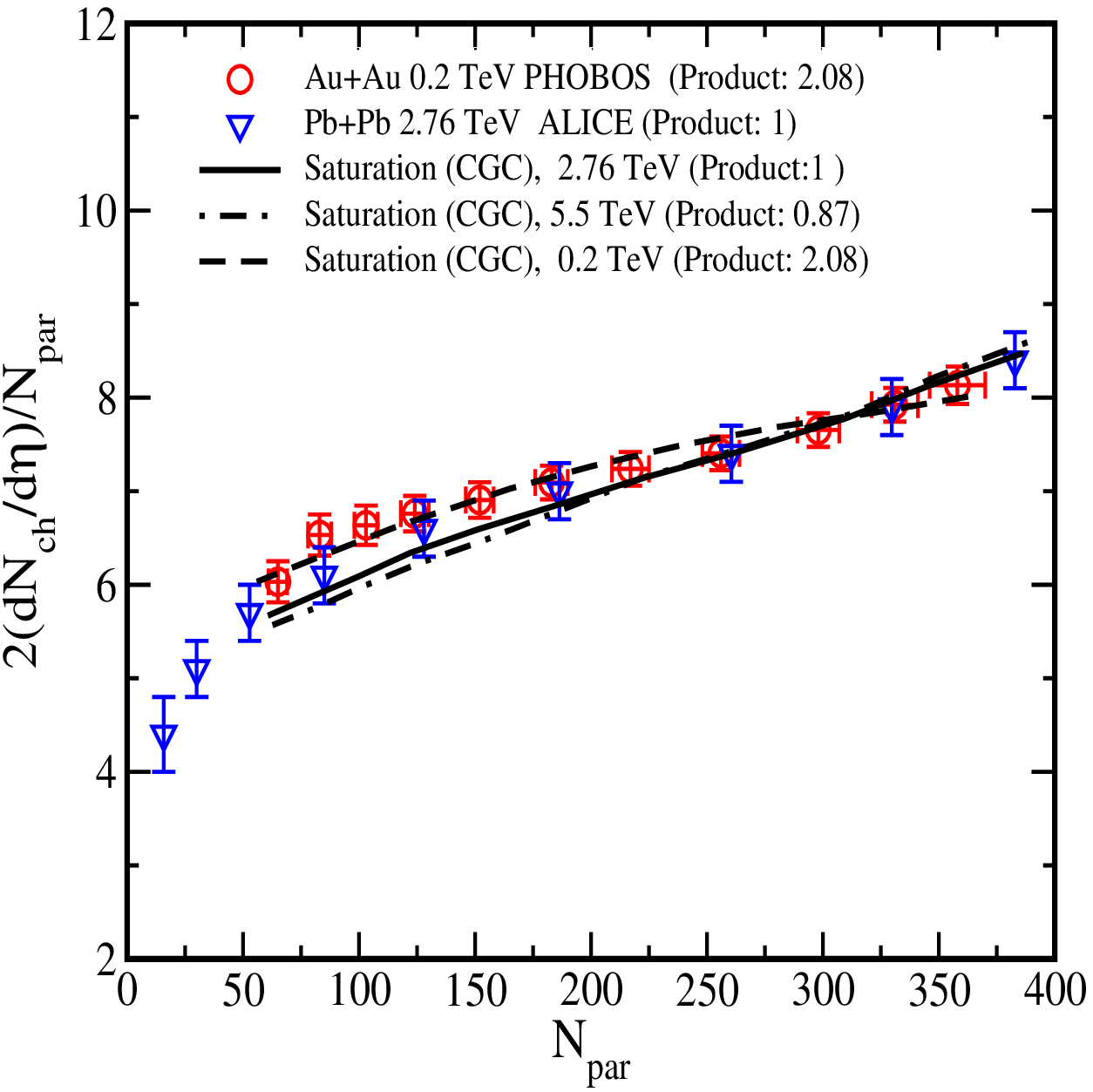}
\caption{Right: The scaled pseudo-rapidity density as a function of number of participant at midrapidity for $AA$ collisions at $0.2, 2.76$ and $5.5$ TeV.  Left: The pseudo-rapidity distribution at RHIC $0.2$ TeV at different centralities.   
Both theoretical predictions and experimental data show gluon
saturation-driven scaling property. We also show in the plots, the
corresponding normalization product factors. The experimental data are from the PHOBOS \cite{rhic1,rhic2} and ALICE \cite{Apb2}
collaboration.  }
\label{scale}
\end{figure}

In \fig{scale} (right), we show the scaled pseudo-rapidity density
$(2/N_{par})(dN_{AA}/d\eta)$ at midrapidity where $N_{par}$ is the
number of participant for a given centrality. The recent ALICE
data \cite{Apb2} at $2.76$ TeV $AA$ collisions reveals interesting scaling
property, namely $(2/N_{par})(dN_{AA}/d\eta)$ at different energies
have the same $N_{par}$ dependence upto a normalization factor. 
In \fig{scale} (right), we show that the saturation results obtained via \eq{kt} for $\sqrt{s}=0.2, 2.76$ and $5.5$ TeV, indeed follow this scaling behaviour. 
 One can observe similar scaling property already
at RHIC \cite{rhic2}, namely $dN_{AA}/d\eta$ at fixed energy but different
centralities falls into a single curve upto a normalization factor, see \fig{scale}
(left). Both scaling properties shown in \fig{scale} can be easily
understood within the CGC picture and follows from simple
\eq{I11}. We expect that the centrality-scaling for $dN_{AA}/d\eta$ at a fixed energy will be also valid at the LHC. Notice that the logarithmic correction due to the
running strong-coupling in the $k_t$ factorization is not shown in the
simple
\eq{I11} and is important. The curves shown in \fig{scale}
are results of full calculation and includes this effect. We predict
that $(2/N_{par})(dN_{AA}/d\eta)$ for $5.5$ TeV $AA$ collisions at
midrapidity to be about $\frac{1}{0.87\pm 0.06}$ times bigger than the corresponding
one at $2.76$ TeV, see \fig{scale} (right).

Finally, one may wonder if a convolution of the fragmentation function in
\eq{kt}, can have the same effect as incorporating the gluon-decay
cascade effect via $\mathcal{N}^{Gluon}_h$. First, one should note
that the main contribution of the $k_t$ factorization for the
multiplicity comes from small $p_T<1.5$ GeV where the fragmentation
functions are not reliable. On the same line, the fragmentation
function is based on a different factorization and QCD evolution
equation and its universality is also questionable in the $k_t$
factorization approach. Moreover, here we were mostly interested to
understand the role of initial-state effects in the hadron productions
in $pp$ and $AA$ collisions.  Therefore, we generalized the
factorization \eq{kt} in order  to incorporate the missing
initial-state effect due to the gluon-decay cascade. We then assumed that the final hadronization is a soft
process and will not change the direction of gluon decays. 
This was also motivated by the fact that the MLLA scheme \cite{LPHD,MDLA} combined
with the LPHD principle provides a good description of data in
$e^+e^-$ and $ep$ collisions upto very small $p_T$ \cite{LPHD}. 
\begin{boldmath}
\newline
\section{Conclusion}
\end{boldmath}
We showed that the basic energy power-law behaviour given in
Eqs.~(\ref{N21}.\ref{N22}) for $pp$ and $AA$ collisions is in
accordance with the saturation/CGC picture. We showed that the
gluon-jet angular-ordering at the decay stage induces extra
energy-dependence in the case that the saturation scale is large. This
contribution has been neglected in previous $k_t$ factorization based
studies. This effect is important for $AA$ collisions where the
saturation scale is larger and gives rise to an extra contribution
about $20-25\%$ to the multiplicity in $AA$ collisions at the
LHC. This explains the observed different energy power-law behaviour
of charged hadron multiplicity in $AA$ and $pp$ collisions at the
LHC. The scaling properties of multiplicity at different energies and
centralities shown in
\fig{scale} and also the energy power-law behaviour of the multiplicity
in $pp$ and $AA$ collisions shown in \fig{NS}, all indicate that the
saturation picture and the CGC scenario provides a unique and efficient
way of describing various experimental data.
   
\begin{figure}[t]
              \includegraphics[width=9cm,height=8cm] {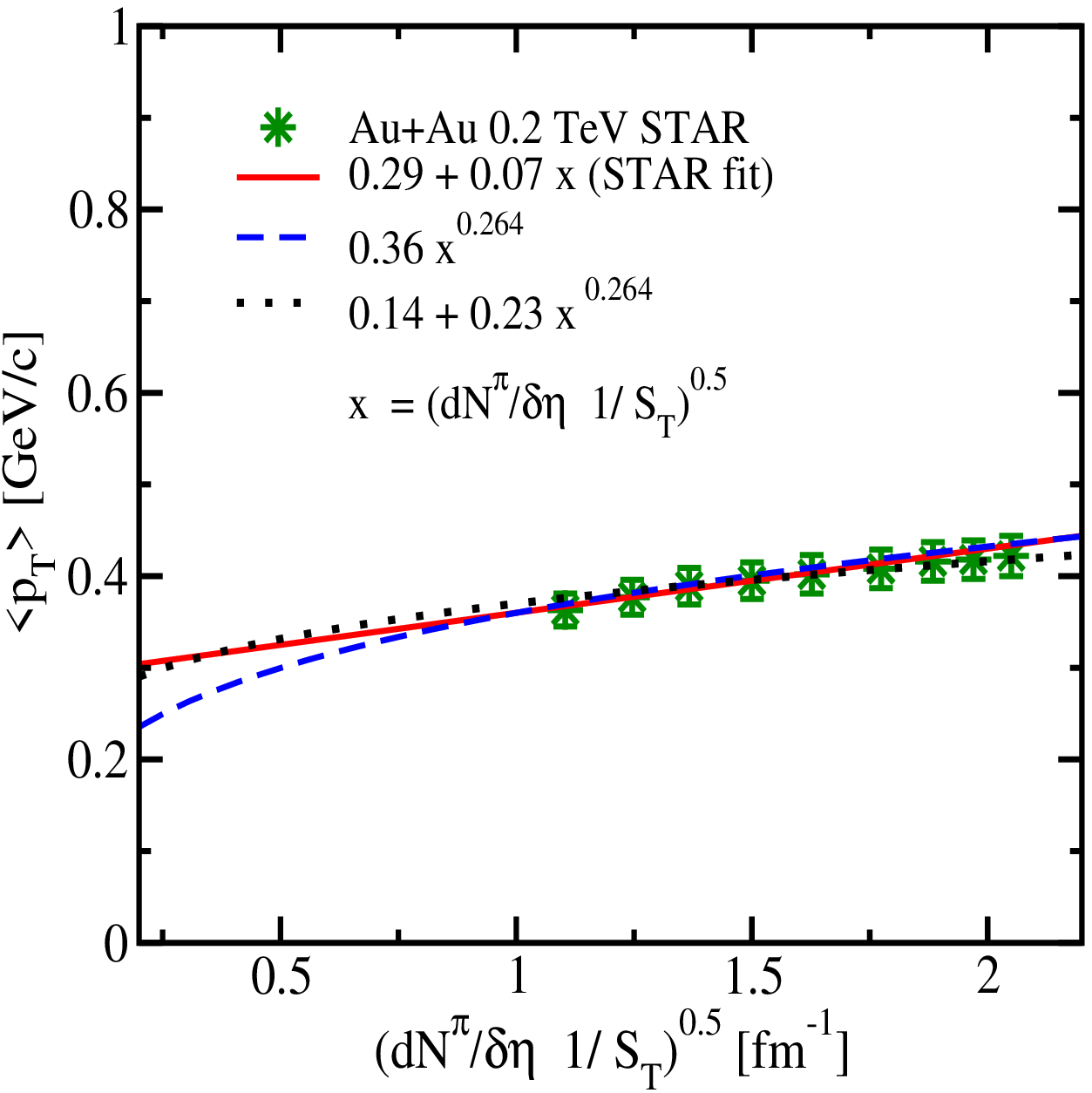}
              \includegraphics[width=7 cm, height=8cm]{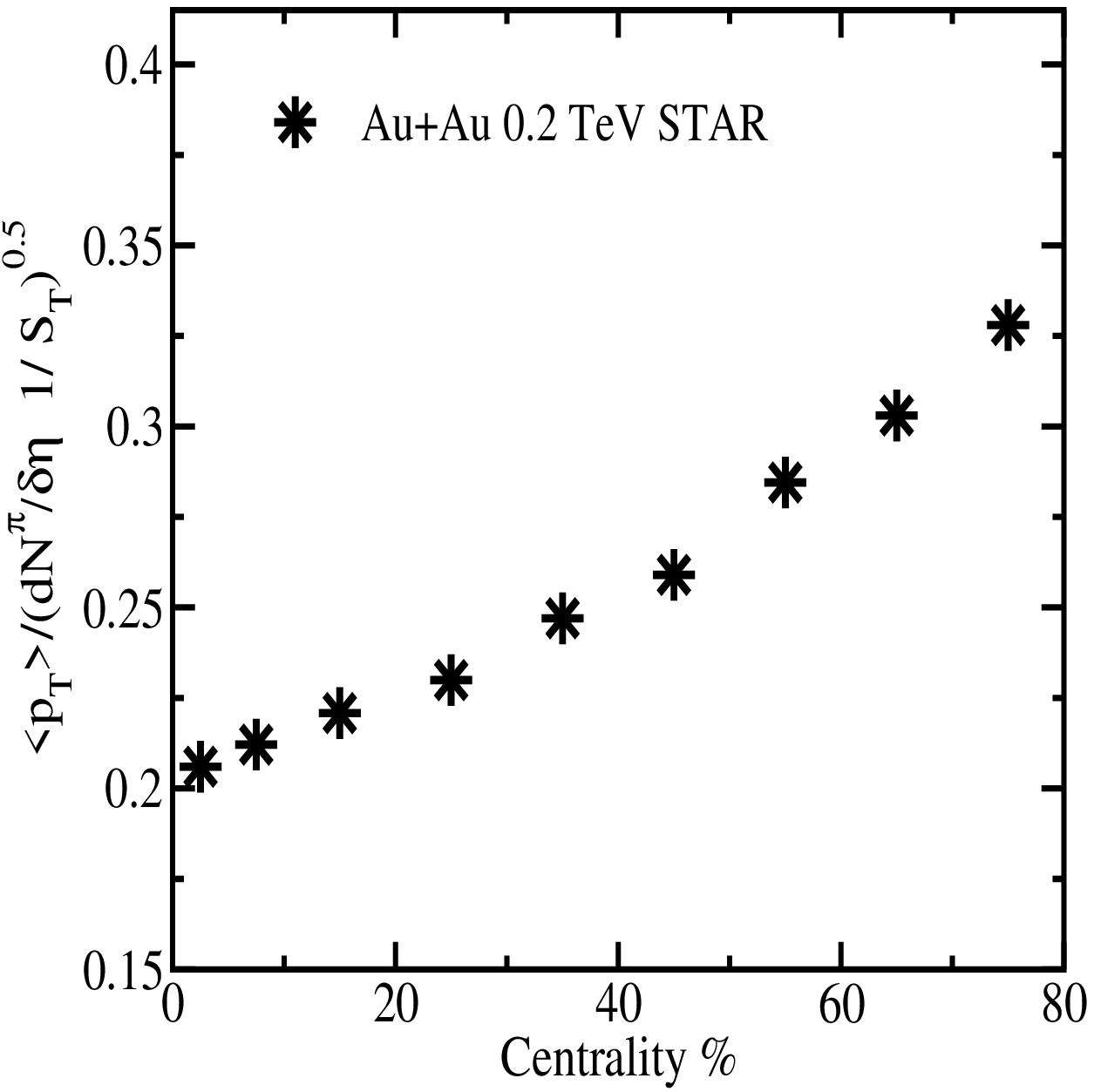}
\caption{Right: The ratio $\langle p_T
\rangle/\sqrt{(dN^{\pi}/d\eta)/S_T}$ at various centralities for Au+Au collisions at $0.2$ TeV. 
The data was constructed from three experimental measurements, the average transverse
momentum $\langle p_T \rangle$ of $\pi^-$, $dN^{\pi}/d\eta$ and $S_T$ \cite{star-f}. Left: Average transverse momenta as a
function of $\sqrt{(dN^{\pi}/d\eta)/\sigma_s}$ for Au+Au collisions at $0.2$
TeV. The experimental data are from the STAR collaboration
\cite{star-f}. }
\label{star}
\end{figure}

After our paper, there has been recently an interesting paper by Lappi
\cite{lap} on the same line.  It is argued by the author of
Ref.~\cite{lap} that the ratio of $\langle p_T
\rangle/\sqrt{(dN/d\eta)/S_T}$ in our approach is in conflict with the
observed experimental data at RHIC. Here we would like to point out
that this is not the case indeed.

It is constructive first to recall the argument of Ref.~\cite{lap}. Let us assume
that based on the LPHD picture one gluon produces $n$ charged hadrons
after fragmentation. Then based on only dimensionality argument we have, 
\bea\label{fff}
 \langle p_T \rangle&\sim & Q_s/n,\nonumber\\
\frac{1}{S_T}\frac{dN}{d\eta}&\sim & nQ_s^2, \\
\eea
where $\langle p_T \rangle$ is the average transverse momentum and
$S_T$ is the overlap area between the colliding nuclei in the
transverse plane.  From above we have $\langle p_T \rangle/\sqrt{(dN/d\eta)/\sigma_s}\sim
\frac{1}{n\sqrt{n}}$. Therefore in our approach, the ratio $\langle p_T \rangle/\sqrt{(dN/d\eta)/\sigma_s}$ decreases for more central collisions in contrast to the KLN type approach \cite{KLN,lap}. 
First, notice that in our approach we have $n\sim N^{Gluon}_{h}$ for
$Q_s > 0.85 \div 1$ GeV corresponding to the excess of charged hadron
production in the presence of jet-decay effects, see
Eqs.~(\ref{I11},\ref{c0}) and \fig{mult}. In
\fig{star} (right) we show the experimental data from the STAR
collaboration \cite{star-f} for $\langle p_T
\rangle/\sqrt{(dN^{\pi}/d\eta)/\sigma_s}$ as a function of centrality. It is seen from \fig{star}
that the ratio $\langle p_T \rangle/\sqrt{(dN/d\eta)/S_T}$ at
$\sqrt{s}=0.2 $ TeV Au+Au collisions decreases for more central
collisions. We expect that based on our model, this ratio should
further suppresses at the LHC in the central $AA$ collisions.

In the KLN type approaches \cite{KLN} we have $\langle p_T \rangle\sim x$ where
$x=\sqrt{(dN/d\eta)/S_T}$ while in our approach we have $\langle p_T
\rangle\sim x^{0.264}$ for the case that the saturation scale is $Q_s>0.85\div 1$ GeV. 
This follows from the fact we have $x\sim Q_s
\sqrt{N^{Gluon}_{h}(Q_s)}$ and $\langle p_T \rangle\sim
Q_s/N^{Gluon}_{h}(Q_s)$ at large saturation scale, see
Eqs.~(\ref{I11},\ref{eee},\ref{c0}). In \fig{star} (left) we show
the average transverse momenta as a function of
$\sqrt{(dN/d\eta)/S_T}$.  The STAR collaboration \cite{star-f} has
found that the experimental data for the charged pion at different
centralities can be described by $\langle p_T
\rangle\approx p_0 + 0.07x$ where the constant $p_0=0.29$ GeV was obtained from a fit and may be interpreted as primordial transverse momentum. 
It seems however that the obtained value of $p_0$ is rather large (and
the coefficient behind $x$ is abnormally small).  Such a rather large value for $p_0$ may be in contradiction
with the notion of asymptotic deconfinement for a dense system, namely, the confinement
radius increases with density \cite{drik,LRAA}. In \fig{star}, we show that a fit
driven by our approach prefers a smaller primordial transverse
momentum of about pion mass $p_0\sim 0.14~\text{GeV}$ (or even
$p_0\sim 0$) and it reasonably describes the same data within the
error bars. Notice that at small multiplicity for very peripheral
collisions our fit and entire saturation formulation is
questionable. 

Finally, we should stress that the jet-decay effects brings rather
small extra contribution which requires a more careful analysis than
only a naive dimensionality argument. We showed in this paper, this
extra contribution is important when the saturation scale is large and
that is in accordance with the existing experimental data in various
reactions. Our main concern in this paper was the combine description
of proton-proton and nuclear data in a contrast to the KLN type
approaches that deal mostly with nuclear reactions.

\begin{boldmath}
\section*{\newline Acknowledgments}
\end{boldmath}
We would like to thank Alex Kovner for useful discussions. 
This work was supported in part by Fondecyt grants 1090312 and 1100648.

\end{document}